\documentclass[conference]{IEEEtran}
\IEEEoverridecommandlockouts
\usepackage{cite}
\usepackage{amsmath,amssymb,amsfonts}
\usepackage{algorithmic}
\usepackage{graphicx}
\usepackage{textcomp}
\usepackage{xcolor}
\def\BibTeX{{\rm B\kern-.05em{\sc i\kern-.025em b}\kern-.08em
    T\kern-.1667em\lower.7ex\hbox{E}\kern-.125emX}}

\usepackage{bm}
\usepackage{algorithm}

\begin{document}

\title{A Massive MIMO Sampling Detection Strategy Based on Denoising Diffusion Model
\thanks{Corresponding author: Zheng Wang (email: wznuaa@gmail.com)}
}

\author{\IEEEauthorblockN{Lanxin He}
\IEEEauthorblockA{\textit{College of Electronic and Information Engineering} \\
\textit{Nanjing University of Aeronautics and Astronautics, Nanjing, China}\\
Email: lanxin\underline{~}he@nuaa.edu.cn}
\and
\IEEEauthorblockN{Zheng Wang and Yongming Huang}
\IEEEauthorblockA{\textit{School of Information Science and Engineering} \\
\textit{Southeast University, Nanjing, China}\\
Email: wznuaa@gmail.com;  huangym@seu.edu.cn}
}
\vspace{-2em}
\maketitle
\vspace{-2em}
\begin{abstract}
The Langevin sampling method relies on an accurate score matching while the existing massive multiple-input multiple output (MIMO) Langevin detection involves an inevitable singular value decomposition (SVD) to calculate the posterior score.
In this work, a massive MIMO sampling detection strategy that leverages the denoising diffusion model is proposed to narrow the gap between the given iterative detector and the maximum likelihood (ML) detection in an SVD-free manner.
Specifically, the proposed score-based sampling detection strategy, denoted as approximate diffusion detection (ADD), is applicable to a wide range of iterative detection methods, and therefore entails a considerable potential in their performance improvement by multiple sampling attempts.
On the other hand, the ADD scheme manages to bypass the channel SVD by introducing a reliable iterative detector to produce a sample from the approximate posterior, so that further Langevin sampling is tractable.
Customized by the conjugated gradient descent algorithm as an instance, the proposed sampling scheme outperforms the existing score-based detector in terms of a better complexity-performance trade-off.

\end{abstract}


\section{Introduction}

The uplink detection in massive multiple-input multiple-output  (MIMO) systems is of vital importance for the sake of realizing the substantial benefits of the evolving MIMO techniques \cite{YangMIMO2015}.
Nevertheless, the computational complexity associated with maximum likelihood (ML) detection is prohibitive for hardware implementation, leading to an urgent request to develop algorithms with more competitive detection trade-off\cite{Gaozhen}.
On the other hand, the score-based generative models, collectively known as \emph{diffusion model}, have lately attracted increasing attention owing to their success in image generation, inpainting, and synthesis\cite{SongSurvey2023}.

Historically, the original diffusion model is proposed in \cite{deepUnsupervised} as an unsupervised generative model.
Inspired by non-equilibrium statistical physics, it gradually destroys the objective distribution through a \emph{forward diffusion} process, and then tries to recover this distribution by a \emph{reverse generative} process.
Later in \cite{DDPM}, a simpler objective function for the denoising diffusion probabilistic model (DDPM) is developed, henceforth simplifying its training and popularizing the diffusion model.
What is more, the equivalence between DDPM and the denoising score matching \cite{SongScore2019} has been shown, which encourages the subsequent unification of DDPM and denoising score matching by stochastic differential equations (SDE) in \cite{SongSDE2020}.
Despite the fact that the diffusion model and the score-based generative model are different in the earlier research, the word ``diffusion model'' now often refers to both of them.

Score matching aims to estimate the \emph{score}, gradient of the logarithm of a density distribution $p(\mathbf{x})$, i.e., $\nabla_\mathbf{x} \log p(\mathbf{x})$.
It captures the characteristic of the objective distribution and is essential for Langevin dynamics sampling \cite{SongScore2019}.
Given $\mathbf{y}$ as an observed variable, in order to apply the Langevin method in massive MIMO detection problem, one has to find out a way to sample from the posterior distribution $p(\mathbf{x} | \mathbf{y})$ instead of $p(\mathbf{x})$.
This can be accomplished by the SNIPS method in \cite{SNIPS}, where a singular value decomposition (SVD) is required to derive the closed-form expression for posterior score.
Afterwards, the first score-based massive MIMO detector\cite{AnnealedZiberstein2022} successfully samples on the posterior distribution using the SNIPS method and manages to implement a list detection.

In this work, we propose another score-based sampling detection method, namely approximate diffusion detection (ADD), as a more flexible and SVD-free scheme.
The main idea is to implement a deterministic detection method stochastically under the denoising diffusion model architecture.
Through multiple sampling attempts, the ADD customized by a particular iterative detector, such as the conjugated gradient descent (CGD), is capable of achieving the near-ML performance.
Assisted by a reliable iterative detection, the sampling on an approximate posterior $\widehat{p}(\mathbf{x} | \mathbf{y})$ characterized by this given detector is tractable.
The technical contribution of this paper is two-fold: 
We propose a score-based sampling detection strategy that can be flexibly applied to a wide range of detection algorithms, while the proposed scheme is able to achieve a more efficient sampling than the existing score-based detection without channel SVD.


\section{Score-Based Massive MIMO Detection}

\subsection{System Model}

For notational simplicity, we consider the real-valued linear system for massive MIMO detection with $K$ transmitting and $N$ receiving antennas as follows
\begin{equation}
	\mathbf{y=Hx+n},
	\label{eqn:real system}
\end{equation}
where the transformation from the complex system model to the real one is straightforward \cite{GPGD}.
We assume the flat Rayleigh fading channel matrix $\mathbf{H} \in \mathbb{R}^{N \times K}$ to be perfectly known, and its entries are independent and identically distributed (i.i.d.) with zero mean and unit variance.
$\mathbf{y} \in \mathbb{R}^{N}$, $\mathbf{x} \in \mathbb{R}^{K}$ and $\mathbf{n} \in \mathbb{R}^{N}$ denote the transmitted signal, the corresponding received signal and the zero-mean additive white Gaussian noise with variance $\sigma_0^2$, respectively.
Denote $\mathcal{Q} = \{ \pm1, \pm3, ..., \pm \sqrt{M}-1 \}$ as the constellation set for $M$-ary quadrature amplitude modulation (QAM).
Given the perfect channel state information (CSI) and an observed $\mathbf{y}$ in (\ref{eqn:real system}), the MIMO detection problem aims to find an estimate of $\mathbf{x}$ that maximizes the posterior probability:
\allowdisplaybreaks{\begin{flalign}
		\widehat{\mathbf{x}}_\mathrm{MAP} &= \underset{\mathbf{x}\in \mathcal{Q}^{K}}{\operatorname{arg~max}} \, p(\mathbf{x} | \mathbf{y}, \mathbf{H}) \notag \\
		&=  \underset{\mathbf{x}\in \mathcal{Q}^{K}}{\operatorname{arg~max}} \, p(\mathbf{y} -  \mathbf{Hx}) p(\mathbf{x}).
\end{flalign}}As a consequence of uniform assumption on prior $p(\mathbf{x})$ and the Gaussian formulation of the noise $\mathbf{n}$, the optimal maximum a posteriori (MAP) solution would reduce to a maximum likelihood (ML) one, namely
\begin{equation}
	\vspace{-.2em}
	\widehat{\mathbf{x}}_\mathrm{ML}=\underset{\mathbf{x}\in \mathcal{Q}^{K}}{\operatorname{arg~min}} \, \|\mathbf{y} - \mathbf{H}\mathbf{x}\|^2.
	\label{eqn:ML Decoding}
		\vspace{-.2em}
\end{equation}


\subsection{Denoising Score Matching}

The \emph{score} of a density distribution $p(\mathbf{x})$ is defined as the gradient of its log-probability, i.e., 
\begin{equation}
	\vspace{-.2em}
	\mathbf{s}(\mathbf{x}) \triangleq \nabla_{\mathbf{x}} \log p(\mathbf{x}).
		\vspace{-.2em}
\end{equation}
The procedure that finds such a function $\mathbf{s}_\theta(\mathbf{x})$ approximating the score is called \emph{score matching}\cite{scoreMatching2005}, where $\theta$ is the parameter either to be fitted in a deep neural network or to be determined by a traditional trace-based method.
However, the trace-based methods are not tractable for large-scale systems, and henceforth the denoising score matching \cite{VincentConnection2011} methodology is leveraged to bypass the trace calculation.

Specifically, the denoising score matching firstly perturbs the objective distribution $p(\mathbf{x})$ with a predefined Gaussian diffusion kernel $q_{\sigma}(\widetilde{\mathbf{x}} \vert \mathbf{x}) = \mathcal{N}(\widetilde{\mathbf{x}}; \mathbf{x}, \sigma^2 \mathbf{I})$, leading to a perturbed distribution $q_\sigma(\widetilde{\mathbf{x}}) = \int p(\mathbf{x}) q_{\sigma}(\widetilde{\mathbf{x}} \vert \mathbf{x}) d\mathbf{x}$.
Here $\sigma$ controls the similarity between the original distribution and the perturbed one.
Then, a score network $\mathbf{s}_\theta(\widetilde{\mathbf{x}})$ is established and optimized to estimate the score of the perturbed distribution $q_\sigma(\widetilde{\mathbf{x}})$ using the following criterion\cite{VincentConnection2011}: 
\begin{equation}
	\theta = \underset{\theta}{\operatorname{arg~min}} \,\, \mathbb{E}_{q_\sigma(\widetilde{\mathbf{x}}\vert \mathbf{x}){p(\mathbf{x})}}[\|\mathbf{s_{\theta}}(\widetilde{\mathbf{x}})-\nabla_{\widetilde{\mathbf{x}}}\log q_\sigma(\widetilde{\mathbf{x}}\vert\mathbf{x})\|^2].
	\label{eqn:denoising SM objective}
\end{equation}
As long as the perturbation is small enough, the optimal network $\mathbf{s}^*_\theta(\mathbf{x})$ satisfies
\begin{equation}
\mathbf{s}^*_\theta(\mathbf{x}) = \nabla_{\mathbf{x}} \log q_\sigma(\mathbf{x}) \approx \nabla_{\mathbf{x}} \log p(\mathbf{x}),
\end{equation}
therefore attaining the score of the objective distribution $p(\mathbf{x})$.

\subsection{Langevin Dynamics for Massive MIMO Detection}

\begin{algorithm}[t]
	\renewcommand{\algorithmicrequire}{\textbf{Input}}
	\renewcommand{\algorithmicensure}{\textbf{Output}}
	\caption{Annealed Langevin Sampling (ALS) Detection}
	\label{alg:ALS}
	\begin{algorithmic}[1]
		\REQUIRE Received signal $\mathbf{y}$, channel matrix $\mathbf{H}$, $\sigma_0$, $T$, $L_A$, noise schedule $\{\sigma_t\}_{t = 1}^T$
		\STATE Calculate the SVD of $\mathbf{H} = \mathbf{U} \bm{\Sigma} \mathbf{V}^T$, $\widetilde{\mathbf{x}}_{T} \sim \mathcal{N}(\mathbf{0}, \sigma_{T}\mathbf{I})$
			\FOR{$t = T, \cdots 1$}
		\STATE Update the step size $\delta_t$ according to \cite{AnnealedZiberstein2022}, $\widehat{\mathbf{x}}_0 = \widetilde{\mathbf{x}}_{t}$
		\FOR{$i = 1, \cdots, L_A$}
		\STATE Calculate the posterior score $\mathbf{s}(\mathbf{x}_{i-1}, \sigma_t)$ in \cite{AnnealedZiberstein2022}
		\STATE $\widehat {\mathbf{x}}_i=\widehat{\mathbf{x}}_{i-1}+\frac{\delta_t}{2}\mathbf{s}(\widehat{\mathbf{x}}_{i-1},\sigma_t) +\sqrt{\delta_t}\mathbf{w}_i$, $\mathbf{w}_i \sim \mathcal{N}(\mathbf{0, I})$
		\ENDFOR
		\STATE $\widetilde{\mathbf{x}}_{t-1} = \widehat{\mathbf{x}}_{L_A}$
		\ENDFOR
		\ENSURE $\widehat{\mathbf{x}} = \widehat{\mathbf{x}}_{L_A}$
	\end{algorithmic}
\end{algorithm}

Once the score $ \nabla_{\mathbf{x}} \log p(\mathbf{x})$ is obtained, the Langevin method \cite{Langevin} is able to produce samples from the density distribution $p(\mathbf{x})$ given the initial $\widehat{\mathbf{x}}_0$ from a prior distribution $\pi$, $\widehat{\mathbf{x}}_0 \sim \pi(\mathbf{x})$, by iterating up to $L$ times as follows,
\begin{equation}
	\widehat{\mathbf{x}}_{i}=\widehat{\mathbf{x}}_{i-1}+\frac{\delta}{2} \nabla_{\mathbf{x}} \log p(\widehat{\mathbf{x}}_{i-1}) +\sqrt{\delta}\mathbf{w}_{i},\,\, i = 1, 2, \cdots, L.
\end{equation}
Here $\mathbf{w}_i \sim \mathcal{N}(\mathbf{0}, \mathbf{I})$, $i$ denotes the sampling time step and $\delta$ is the sampling step size.
However, the naive application of Langevin method would encounter irregular fluctuation, estimation inaccuracy as well as slow mixing problem.
Henceforth the \emph{annealed Langevin} strategy that adopts multiple noise levels is preferred to obviate these difficulties\cite{SongScore2019}.
Specifically, a noise schedule $\{\sigma_t\}_{t = 1}^{T}$ that satisfies $\dfrac{\sigma_1}{\sigma_2}=\cdots=\dfrac{\sigma_{T-1}}{\sigma_T}<1$ divides the whole perturbation into several intermediate \emph{forward diffusion} processes $q_{\sigma_t}(\widetilde{\mathbf{x}}_{t}|\mathbf{x}) \sim \mathcal{N}(\widetilde{\mathbf{x}}_{t}; \mathbf{x}, \sigma_t^2 \mathbf{I})$.
This can be re-parameterized as the following Markov chain:
\begin{equation}
	\widetilde{\mathbf{x}}_t=\widetilde{\mathbf{x}}_{t-1}+\sqrt{\sigma_t^2-\sigma_{t-1}^2}\mathbf{z}_{t},\quad t=1,\cdots,T,
	\label{eqn:forward Markov chain}
\end{equation}
where $\mathbf{z}_t \sim \mathcal{N}(\mathbf{0, I})$ is a Gaussian random noise and $\widetilde{\mathbf{x}}_0 \triangleq \mathbf{x}$.

Then, an effort to recover these perturbed $\{\widetilde{\mathbf{x}}_t\}_{t=1}^{T}$ reversely is made in the \emph{reverse generative} process, where a noise conditional score network $\mathbf{s}_{\theta}(\mathbf{x}, \sigma_t)$ is involved to train.
Once the training is done, a successive $L_A$-times sampling is conducted along $q_{\sigma_t}(\widetilde{\mathbf{x}}_t)$ in a descending order, i.e., $\widetilde{\mathbf{x}}_T, \cdots, \widetilde{\mathbf{x}}_1$.
To be more specific, for each $L_A$-times sampling, it starts from $\widehat{\mathbf{x}}_0 = \widetilde{\mathbf{x}}_t$, and the final sample $\widehat{\mathbf{x}}_{L_A}$ is treated as an estimate of perturbed variable for the next iteration, namely $\widetilde{\mathbf{x}}_{t-1} = \widehat{\mathbf{x}}_{L_A}$, as follows,
\begin{equation}
	\widehat {\mathbf{x}}_i=\widehat{\mathbf{x}}_{i-1}+\frac{\delta_t}{2}\mathbf{s}_{\theta}(\widehat{\mathbf{x}}_{i-1},\sigma_t) +\sqrt{\delta_t}\mathbf{w}_i,  i = 1, 2, \cdots, L_A.
\end{equation}
Here $\delta_t = \varepsilon \cdot \sigma_t^2 / \sigma_T^2$, with $\varepsilon$ being the annealed learning rate, updates corresponding to $\sigma_t$ in a reverse order, i.e., from $\sigma_T$ to $\sigma_1$.
Clearly, this annealed Langevin strategy requires $L_A \times T$ times sampling in total to produce a single reliable sample, and generally $L_A \times T \leq L$ stands as a result of the improved mixing rate of annealed Langevin dynamics.

\begin{figure*}[h]
	\hspace{-1em}\centerline{\includegraphics[width=1.05\textwidth]{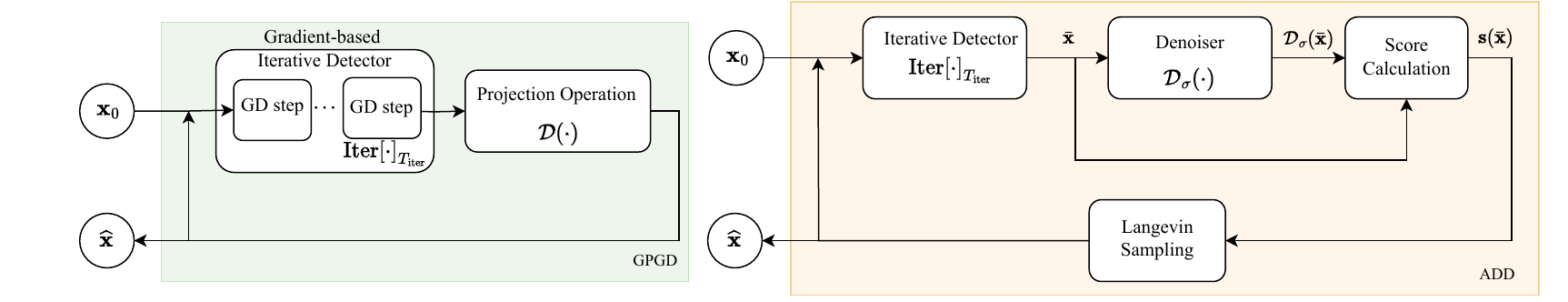}}
	\vspace{-.5em}
	\caption{Framework of GPGD and the proposed ADD sampling scheme.}
	\label{fig:framework}
\end{figure*}

Nevertheless, the detection problem needs to tackle the issue of sampling on the posterior density distribution $p(\mathbf{x} \vert \mathbf{y})$ rather than $p(\mathbf{x})$.
One might handle this by applying the Bayes' rule and rewriting the posterior score as
\begin{equation}
	\nabla_{\widehat{\mathbf{x}}_t}\log p(\widehat{\mathbf{x}}_t\vert \mathbf{y})=\nabla_{\widehat{\mathbf{x}}_t}\log p(\widehat{\mathbf{x}}_t)+\nabla_{\widehat{\mathbf{x}}_t}\log p(\mathbf{y}\vert\widehat{\mathbf{x}}_t),
\end{equation}
whereas a singular value decomposition (SVD) of the channel matrix $\mathbf{H}$ is required to formulate this posterior score explicitly, thereby imposing strains on computation complexity for large-scale systems. 
This annealed Langevin sampling (ALS) detection \cite{AnnealedZiberstein2022} is outlined in Alg.\ref{alg:ALS} for one single sampling trajectory.
Finally, the entire Langevin dynamics produces an $S$-length sample list $\mathcal{L} = \{\widehat{\mathbf{x}}^{(1)}, \widehat{\mathbf{x}}^{(2)}, \cdots, \widehat{\mathbf{x}}^{(S)} | \widehat{\mathbf{x}} \sim p(\mathbf{x} | \mathbf{y})\}$ for detection, amongst which the sample that minimizes following Euclidean distance is distinguished as the final estimate:
\begin{equation}
	\mathbf{\widehat{x}}= \underset{\mathbf{x} \in \mathcal{L}}{\operatorname{arg~min}} \,\, \|\mathbf{y}-\mathbf{H}\mathbf{x}\|^2.
	\label{eqn:list detection}
\end{equation}


\section{Approximate Diffusion Detection Strategy}

In this section, we present the proposed ADD methodology in detail.
We mention that the neural network training is not necessary here, but ADD is still amenable to further deep generative detection network extension.
Meanwhile, it is shown by \cite{SongSDE2020} that the Markov chain in (\ref{eqn:forward Markov chain}) is equivalent to the following SDE:
\begin{equation}
	\mathrm{d}\mathbf{x}=\sqrt{\frac{\mathrm{d}\left[\sigma^2(t)\right]}{\mathrm{d}t}}\mathrm{d}\bm{\omega},
	\label{eqn:SDE forward}
\end{equation}
where $\bm{\omega}$ is the standard Wiener process, and its diffusion coefficient is $g(t) = \sqrt{\frac{\mathrm{d}\left[\sigma^2(t)\right]}{\mathrm{d}t}}$.
Basically, this SDE assumes that, as $T \rightarrow \infty$, the Markov chain $\{\widetilde{\mathbf{x}}_t\}_{t = 1}^{T}$ turns into a continuous stochastic process $\{\widetilde{\mathbf{x}}(t)\}_{t=0}^1$, where the integer index $t = 1, \cdots, T$ is substituted by a continuous time variable $t \in [0, 1]$.
For the same reason, the noise schedule $\{\sigma_t\}_{t = 1}^{T}$ alters to a function $\sigma(t)$ and $\mathbf{z}_t$ becomes a Gaussian process $\mathbf{z}(t)$.
Accordingly, its reverse SDE is given by
\begin{equation}	\mathrm{d}\mathbf{x}=-g(t)^2\nabla_{\mathbf{x}} \log p_t(\mathbf{x})\mathrm{d}t+g(t)\mathrm{d}\bar{\bm{\omega}},
	\label{eqn:SDE reverse}
\end{equation}
where $\bar{\bm{\omega}}$ is the Wiener process through a reverse time and $\nabla_{\mathbf{x}} \log p_t(\mathbf{x})$ is the score at time $t$.
The SDE-manner formulation unifies both the denoising score matching and the diffusion model, and therefore is adopted here for further convenience.
Note that for the forward process in (\ref{eqn:SDE forward}), the time variable $t$ evolves from $t = 0$ to $t = 1$.
By contrast, it reduces from $t = 1$ to $t = 0$ in the reverse process of (\ref{eqn:SDE reverse}), which accounts for distinguishing this reverse generating from the forward diffusion process.

%

\subsection{Framework}

The proposed ADD strategy is inspired by the generalized projected gradient descent (GPGD) method in \cite{GPGD}, where a projection operator alternates with an iterative gradient-based detector, but the proposed strategy is also applicable to other reliable detection algorithms.
Besides, the ADD manages to circumvent the calculation of posterior score by SVD and results in a computational efficiency.

Typically, with respect to the GPGD method, a gradient-based iterative detector, denoted by $\mathrm{Iter}\left[ \cdot\right]_{T_\mathrm{iter}}$, implements $T_\mathrm{iter}$ times gradient descent (GD) steps to provide an estimate of the signal.
Afterwards a projection operator $\mathcal{D}$ tries to find the ideal counterpart of this estimate in a specific discrete domain.
One iteration of this process can be expressed as
\begin{equation}
	\mathbf{x}_{i+1} = \mathcal{D}\left( \mathrm{Iter}\left[ \mathbf{x}_{i}, \mathbf{y};  \mathbf{H}\right]_{T_\mathrm{iter}} \right),
\end{equation}
with $i$ being the iteration index.
In practical, the projection operator $\mathcal{D}$ might be constructed by a deep neural network and trained under a minimum mean square error (MSE) criterion, but finding a good enough projection is always time-consuming and requires dedicated network design.
The main idea behind the proposed strategy is to transform the deterministic algorithm in a stochastic manner, where the performance gain can be attained by multiple sampling attempts, thus shrinking the gap to the optimal ML detection.

In particular, during the ADD, an iterative method is utilized to produce a relatively reliable solution $\bar{\mathbf{x}} = \mathrm{Iter}[\mathbf{x}]_ {T_\mathrm{iter}}$ that supports the subsequent score estimation.
This variable characterizes the distribution of an approximate posterior, denoted by $\widehat{p}(\mathbf{x} | \mathbf{y})$, which is the key to eliminate the closed-form score calculation.
Based on this, a denoiser $\mathcal{D}_{\sigma}(\cdot)$ parameterized by $\sigma$, plays a similar role as the projection operator in GPGD method to give a minimum MSE estimate $\mathcal{D}_{\sigma}(\bar{\mathbf{x}})$.
By doing so, the score of approximate posterior can be given by the Tweedie’s identity, whose feasibility can be found in \cite{ZahraImplicit2021}:
\begin{equation}
	\mathbf{s}(\bar{\mathbf{x}}) = \frac{\mathcal{D}(\bar{\mathbf{x}}) - \bar{\mathbf{x}}}{\sigma^2}.
	\label{eqn:score}
\end{equation}
Leveraging this score, the Langevin sampling can be conducted to sample from the approximate posterior $\widehat{p}(\mathbf{x} | \mathbf{y})$.
The proposed ADD strategy and the GPGD method are compared in Fig.\ref{fig:framework}, where their related structures can be observed.
Since the denoiser in ADD works as an MSE minimizer, which is often the role played by the projection in GPGD method, extra projection in the ADD is not necessary.

\subsection{Implementation}

\begin{algorithm}[t]
	\renewcommand{\algorithmicrequire}{\textbf{Input}}
	\renewcommand{\algorithmicensure}{\textbf{Output}}
	\caption{Approximate Diffusion Detection (ADD)}
	\label{alg:ADD}
	\begin{algorithmic}[1]
		\REQUIRE Received signal $\mathbf{y}$, channel matrix $\mathbf{H}$, symbol energy $E_s$, $\sigma_0$, $\sigma_{\min}$, $\sigma_{\max}$, $\epsilon$, $T$, $T_\mathrm{iter}$, $S$
		
		\STATE Calculate the noise schedule $\sigma(t)= \sigma_{\min}\left(\frac{\sigma_{\max}}{\sigma_{\min}}\right)^t$ and the diffusion coefficient $g(t)=\sigma_{\min}\left(\frac{\sigma_{\max}}{\sigma_{\min}}\right)^t\sqrt{2\log\frac{\sigma_{\max}}{\sigma_{\min}}}$ for $t\in\left[\epsilon,1\right]$ with the interval $\Delta t = \frac{1 - \epsilon}{T}$
		\FOR{$j = 1, ..., S$}
		\STATE $t = 1$, $\widetilde{\mathbf{y}}(0) = \mathbf{y}$,
		$\widehat{\mathbf{x}}(1) \sim \mathcal{N}(\mathbf{0}, \sigma_{\max}^2 \mathbf{I})$
		\WHILE{$t \neq \epsilon$}
		
		\STATE Sample $\widetilde{\mathbf{y}}(t) \sim \mathcal{N}\left(\widetilde{\mathbf{y}}(t);\mathbf{y}(0),\sigma(t)^2\mathbf{H}\mathbf{H}^T\right)$
		\STATE $t = t - \Delta t$
		\STATE Evoke Function \ref{alg:CGD} to get $\bar{\mathbf{x}}(t)$
		\STATE Estimate the score $\mathbf{s}\big( \bar{\mathbf{x}}(t) \big) = \dfrac{\mathcal{D}_{\sigma(t)}(\bar{\mathbf{x}}(t)) - \bar{\mathbf{x}}(t)}{\sigma(t)^2}$, where $\mathcal{D}_{\sigma(t)}(\cdot)$ adopts the LGD form in (\ref{eqn:LGD})
		\STATE Sample $\mathbf{w}(t) \sim \mathcal{N}(\mathbf{0}, \mathbf{I})$ and get
		\begin{equation}
			\widehat{\mathbf{x}}(t) = \bar{\mathbf{x}}(t) - g^2(t)\cdot \mathbf{s}\big( \bar{\mathbf{x}}(t) \big) \Delta t + g(t) \sqrt{\Delta t}\cdot \mathbf{w}(t) \notag
		\end{equation}
		\ENDWHILE
		\STATE Include $\widehat{\mathbf{x}}(\epsilon)$ to the list $\mathcal{L}$ of length $S$.
		\ENDFOR
		\ENSURE $\widehat{\mathbf{x}} = \underset{\mathbf{x} \in \mathcal{L}}{\operatorname{arg~min}} \,\, \|\mathbf{y}-\mathbf{H}\mathbf{x}\|^2.$
	\end{algorithmic}
\end{algorithm}

The implementation details of the proposed ADD method are elaborated in the following.
Here we define the noise schedule function as
\begin{equation}
	\sigma(t)=\sigma_{\min}\left(\frac{\sigma_{\max}}{\sigma_{\min}}\right)^t,\quad t\in\left[0,1\right],
\end{equation}
with $\sigma(0) = \sigma_{\min} \triangleq \sigma_1$ and $\sigma(1) = \sigma_{\max} \triangleq \sigma_T$.
We still denote $T$ as the total sampling times, and henceforth the interval between each individual time step is $\Delta t = \frac{1}{T}$.

In this way, the one-step transition probability of the forward diffusion process can be written as
\allowdisplaybreaks{\begin{flalign}
&p\left(\widetilde{\mathbf{x}}(t)\vert\widetilde{\mathbf{x}}(t-\Delta t)\right) \notag\\
&\hspace{-.2em}=\hspace{-.1em}\mathcal{N}\hspace{-.2em}\left(\widetilde{\mathbf{x}}(t);\widetilde{\mathbf{x}}(t \! -\!\Delta t),\!\left(\sigma^2(t)\!-\! \sigma^2(t-\Delta t)\right) \! \mathbf{I}\right),t\!\in\![0,1],
\end{flalign}}and the corresponding $t$-step counterpart is 
\begin{equation}
	p\left(\widetilde{\mathbf{x}}(t)\vert\mathbf{x}(0)\right)=\mathcal{N}\left(\widetilde{\mathbf{x}}(t);\mathbf{x}(0),\sigma(t)^2\mathbf{I}\right).
	\label{eqn:t-step transition}
\end{equation}
According to this, the perturbed variable $\widetilde{\mathbf{x}}(t)$ in the forward diffusion is re-parametrized as $\widetilde{\mathbf{x}}(t) = \mathbf{x}(0) + \sigma(t)\mathbf{z}(t)$.
Therefore, the perturbed received signal at time $t$ can be formulated as
\allowdisplaybreaks{\begin{flalign}
	\widetilde{\mathbf{y}}(t)&=\mathbf{H}\widetilde{\mathbf{x}}(t)+\mathbf{n}=\mathbf{H}\big(\mathbf{x}(0)+\sigma(t)\mathbf{z}(t)\big)+\mathbf{n}\notag\\
	&=\mathbf{H}\mathbf{x}(0)+\mathbf{n}+\sigma(t)\mathbf{H}\mathbf{z}(t)=\mathbf{y}+\sigma(t)\mathbf{H}\mathbf{z}(t)
\end{flalign}}with $\mathbf{x}(0) = \mathbf{x}$.
Now the $t$-step transition probability of $\widetilde{\mathbf{y}}(t)$ is also attainable:
\begin{equation}
	p\left(\widetilde{\mathbf{y}}(t)\mid\mathbf{y}(0)\right)\sim\mathcal{N}\left(\widetilde{\mathbf{y}}(t);\mathbf{y}(0),\sigma(t)^2\mathbf{H}\mathbf{H}^T\right),
	\end{equation}
where $\mathbf{y}(0) = \mathbf{y}$.
In this way, sampling on the perturbed received signal $\widetilde{\mathbf{y}}(t)$ is tractable.
Hence, starting from the time $t=1$, the reverse sample $\widehat{\mathbf{x}}(t)$ and forward $\widetilde{\mathbf{y}}(t)$ can cooperate to give an estimate $\bar{\mathbf{x}}(t)$ by the iterative detection method, i.e.,
\begin{equation}
	\bar{\mathbf{x}}(t) = \mathrm{Iter}\left[\widehat{\mathbf{x}}(t), \widetilde{\mathbf{y}}(t); \mathbf{H}\right]_ {T_\mathrm{iter}}.
\end{equation}
This estimate is viewed as a sample from the approximate posterior $\widehat{p}(\mathbf{x} | \mathbf{y})$ depending on the chosen iterative detection method.

Now that the variable $\bar{\mathbf{x}}$ encapsulates information about $\mathbf{y}$ and $\mathbf{H}$, once a minimum MSE estimate of $\bar{\mathbf{x}}$ is obtained, the score of $\widehat{p}(\mathbf{x} | \mathbf{y})$ can be analyzed according to (\ref{eqn:score}): $	\mathbf{s}\big(\bar{\mathbf{x}}(t)\big) =\dfrac{\mathcal{D}_{\sigma(t)}(\bar{\mathbf{x}}(t)) - \bar{\mathbf{x}}(t)}{\sigma(t)^2}$.
The required denoiser $\mathcal{D}_{\sigma(t)}$ can be established by a deep neural network, while for now we directly use the training-free estimator as in \cite{AnnealedZiberstein2022}.
Specifically, this estimator evaluates every component $x_k$ of the input $\mathbf{x}$ aided by the one-dimensional (1-D) lattice Gaussian distribution (LGD), namely
\begin{equation}
\!\! p_{\mathcal{Q}}(x_k=\widehat{x}_k; \bar{x}_k, \sigma)\!\triangleq\! \frac{1}{Z_\mathcal{Q}} \! \exp \! \left(\! \frac{-\|\widehat{x}_k-\bar{x}_k\|^2}{2\sigma^2}\!\right)\!, \widehat{x}_k \in \mathcal{Q},\!	\label{eqn:LGD}
\end{equation}
with $Z_{\mathcal{Q}}=\sum_{\widehat{x}_{k}\in \mathcal{Q}}\exp\left(\frac{-\| \widehat{x}_{k}-\bar{x}_k||^{2}}{2\sigma^{2}}\right)$ a normalization scalar.

\begin{figure}[t]
	\vspace{-.2em}\centerline{\includegraphics[width=0.43\textwidth]{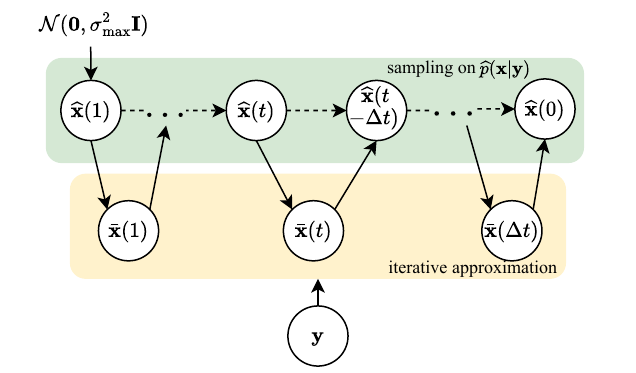}}
	\caption{The reverse generative process in the proposed ADD strategy.}
	\label{fig:generative process}
	\vspace{-1em}
\end{figure}

To this end, we can generate a new sample $\widehat{\mathbf{x}}$ utilizing the attained score as (\ref{eqn:SDE reverse}) indicates, leading to
\begin{equation}
\widehat{\mathbf{x}}(t) = \bar{\mathbf{x}}(t) - g^2(t)\cdot \mathbf{s}\big( \bar{\mathbf{x}}(t) \big) \Delta t + g(t) \sqrt{\Delta t}\cdot \mathbf{w}(t),
\end{equation}
where $\mathbf{w}(t)$ is the Gaussian process from $\mathbf{w}_t$.
As the sampling goes on from $t=1$ to $t=0$, the final sample $\widehat{\mathbf{x}}(0)$ is treated as sampled from the approximate posterior distribution $\widehat{p}(\mathbf{x} | \mathbf{y})$.
The whole generative procedure is illustrated in Fig.\ref{fig:generative process}.
This completes a single trajectory for the sampling, and this procedure continues up to $S$ times, generating a candidate list for final decision.
The overall algorithm is outlined in Alg.\ref{alg:ADD}, where the CGD method\cite{NumericalOptimization}, as presented in \textbf{Func.\ref{alg:CGD}}, is chosen to customize the $\mathrm{Iter}\left[\cdot\right]_ {T_\mathrm{iter}}$ procedure as a simple instance.
It is suggested that $\sigma_{\min}$ should be set as a very small value, like $\sigma_{\min} = 0.01$, and $\sigma_{\max}$ close to the symbol energy $E_s$.
Meanwhile, a extremely small-valued $\epsilon > 0$ is recommended for numerical stability.

\begin{algorithm}[t]
	\floatname{algorithm}{Function}
	\setcounter{algorithm}{0}
	\renewcommand{\algorithmicrequire}{\textbf{Input}}
	\renewcommand{\algorithmicensure}{\textbf{Return}}
	\caption{Conjugate Gradient Descent (CGD)}
	\label{alg:CGD}
	\begin{algorithmic}[1]
		\REQUIRE $\mathbf{y} = \widetilde{\mathbf{y}}(t)$, $\mathbf{x}_0 = \widehat{\mathbf{x}}(t)$, $\mathbf{H}$, $T_\mathrm{iter}$
		\STATE Initialization: $\mathbf{A} = \mathbf{H}^T\mathbf{H} + \sigma_0 E_s\mathbf{I}$, $\mathbf{b} = \mathbf{H}^T \mathbf{y}$\\$\mathbf{r}_0 = \mathbf{A}\mathbf{x}_0 - \mathbf{b}$, $\mathbf{d}_0 = -\mathbf{r}$
		\FOR{$i = 0, 1, \cdots, T_\mathrm{iter}-1$}
		\STATE $\alpha_{i + 1} = ({\mathbf{r}_i^T \mathbf{r}_i}) / ({\mathbf{r}_i^T \mathbf{A} \mathbf{d}_i}$)
		\STATE $\mathbf{x}_{i+1} = \mathbf{x}_i + \alpha_{i+1} \mathbf{d}_i$
		\STATE $\mathbf{r}_{i+1} = \mathbf{r}_i  + \alpha_{i+1} \mathbf{Ad}_i$
		\STATE $\beta_{i+1} = (\mathbf{r}_{i+1}^T \mathbf{r}_{i+1}) / (\mathbf{r}_i^T \mathbf{r}_i)$
		\STATE $\mathbf{d}_{i+1} = -\mathbf{r}_{i+1} + \beta_{i+1} \mathbf{d}_{i}$
		\ENDFOR
		\ENSURE $\mathbf{x}_{T_\mathrm{iter}}$
	\end{algorithmic}
\end{algorithm}

\subsection{Complexity Analysis}

For one single trajectory, the computation for ADD strategy mainly consists of three parts: selected iterative detector, score calculation and sampling.
Particularly, with respect to the sampling, the generation of $\widetilde{\mathbf{y}}(t) \sim \mathcal{N}\left(\widetilde{\mathbf{y}}(t);\mathbf{y}(0),\sigma(t)^2\mathbf{H}\mathbf{H}^T\right)$ is achieved by re-parameterizing $\widetilde{\mathbf{y}}(t) = \mathbf{y}+\sigma(t)\mathbf{H}\mathbf{z}(t)$, which involves multiplying $\mathbf{H}$ to a noise vector $\mathbf{z}(t)$ with $NK$ multiplications.
Besides, there are $2T$ times random Gaussian noise generation, one $\mathbf{z}(t)$ and one $\mathbf{w}(t)$ for each iteration, but their complexity can be neglected.
Moreover, since the denoiser $\mathcal{D}_{\sigma}(\cdot)$ adopted in this work evaluates by 1-D LGD, whose complexity is $O(MK)$ for $M$-QAM, the score calculation part does not require any neural network computing.
Finally, as for the iterative detector, taking CGD as an example, its complexity is dominated by calculation of $\mathbf{H}^T\mathbf{H}$, which is of order $O(NK^2)$, and the complexity for one single iterative step is $O(K^2)$.
Therefore, the overall complexity of the proposed ADD strategy is $O(NK^2 + T(NK + T_{\mathrm{iter}}K^2 + MK))$ in a polynomial manner.
Compared to the ALS method, whose complexity is $O(NK^2 + L_A T (K^2 + MK))$\cite{AnnealedZiberstein2022} and largely affected by the SVD operation, the dominant computation of ADD is the chosen iterative detector.
Here for the CGD case, the calculation of Hermitian matrix $\mathbf{H}^T\mathbf{H}$ is much simpler than SVD computation, and additionally ADD provides a more flexible scheme with a substitutable iterative detector, leading to a possibility in future complexity-reduction.

\subsection{Discussions On Deep Learning Extension}
In the following, we show that further deep learning extension is possible in the proposed ADD scheme.
Consider the diffusion kernel $q_\sigma(\widetilde{\mathbf{x}}\vert \mathbf{x})=\mathcal{N}(\widetilde{\mathbf{x}}; \mathbf{x},\sigma^2I)$ in the mentioned denoising score matching.
It is easy to derive that its score has the following form:
\begin{equation}
	\nabla_{\widetilde{\mathbf{x}}}\log q_\sigma(\widetilde{\mathbf{x}}\mid\mathbf{x})=-(\widetilde{\mathbf{x}}-\mathbf{x})/\sigma^2.
	\label{eqn:gradient log}
\end{equation}
According to the objective function in (\ref{eqn:denoising SM objective}), the training of a denoising score network $\mathbf{s}_{{\theta}}(\widetilde{\mathbf{x}},\sigma)$ turns to minimize
\begin{equation}
	\!\!\ell({\theta};\sigma) = \mathbb{E}_{p(\mathbf{x})}\mathbb{E}_{\widetilde{\mathbf{x}}\sim q_\sigma(\widetilde{\mathbf{x}}\vert \mathbf{x}) }\left[\left\|\mathbf{s}_{{\theta}}(\widetilde{\mathbf{x}},\sigma)+\frac{\widetilde{\mathbf{x}}-\mathbf{x}}{\sigma^2}\right\|^2\right].
	\label{eqn:NCSN loss}
\end{equation}
Again, $\widetilde{\mathbf{x}}$ can be attained by the re-parameterization $\widetilde{\mathbf{x}} = \mathbf{x} + \sigma \mathbf{z}$ with $\mathbf{z} \sim \mathcal{N}(\mathbf{0, I})$.
Henceforth, by substituting $(\widetilde{\mathbf{x}} - \mathbf{x}) / \sigma = \mathbf{z}$, the  loss function in (\ref{eqn:NCSN loss}) can be transformed into 
\begin{equation}
	\ell({\theta};\sigma) = \mathbb{E}_{p(\mathbf{x})} \mathbb{E}_{\mathbf{z} \sim \mathcal{N}(\mathbf{0, I})} \left[\left\|\mathbf{s}_{{\theta}}(\widetilde{\mathbf{x}},\sigma)+\frac{\mathbf{z}}{\sigma}\right\|^2\right].
\end{equation}
This corresponds to the \emph{unsupervised} training scheme used in diffusion generative model.

With the Tweedie’s identity, the score of a Gaussian perturbed variable is estimated by a minimum MSE denoiser $\mathcal{D}_{\sigma}(\cdot)$ to get $\mathbf{s}(\widetilde{\mathbf{x}}, \sigma) = \dfrac{\mathcal{D}_\sigma(\widetilde{\mathbf{x}}) - \widetilde{\mathbf{x}}}{\sigma^2}$, as in the proposed ADD scheme.
Now if we directly use this score to replace the trainable score network $\mathbf{s}_{\theta}(\widetilde{\mathbf{x}}, \sigma)$ in (\ref{eqn:NCSN loss}), the loss function now can be estimated as
\begin{equation}
\ell({\theta};\sigma) = \mathbb{E}_{p(\mathbf{x})} \mathbb{E}_{\widetilde{\mathbf{x}}\sim q_\sigma(\widetilde{\mathbf{x}}\vert \mathbf{x}) } \left[\left\| \mathcal{D}_{\theta, \sigma}(\widetilde{\mathbf{x}})- \mathbf{x}\right\|^2\right],
\label{eqn:ADDnet loss}
\end{equation}
where a subscript $\theta$ that represents the trainable parameter is added to the denoiser.
By doing so, the problem of training a score network $\mathbf{s}_\theta(\widetilde{\mathbf{x}}, \sigma)$ becomes finding a denoising network $\mathcal{D}_{\theta, \sigma}(\widetilde{\mathbf{x}})$ that minimizes the loss function (\ref{eqn:ADDnet loss}), which turns the unsupervised training into a \emph{supervised} one.
The labels are original data $\mathbf{x}$ and an ideal $\mathcal{D}_{\theta, \sigma}(\widetilde{\mathbf{x}})$ shall be able to recover $\mathbf{x}$ given the perturbed data $\widetilde{\mathbf{x}}$ diffused by $q_{\sigma}(\widetilde{\mathbf{x}} | \mathbf{x})$.
Some inspirations might be found in the denoising auto-encoder (DAE) architecture \cite{DAE}.
On the other hand, based on (\ref{eqn:ADDnet loss}), the 1-D LGD denoiser used in ADD inevitably results in a considerable quantization error in the beginning of sampling, where the noise setting $\sigma(t)$ is relatively large.
Henceforth, there is a potential performance gain in the ADD scheme to be realized by means of deep learning, leading to a deep generative detection network.
%

\section{Simulations}

This section examines the performance of ADD strategy, where the CGD is selected to be the inner iterative detector and perfect CSI is assumed at the receiver.
For comparison purposes, the performances of ALS, minimum mean square error (MMSE) , MMSE-based successive interference cancellation (MMSE-SIC) and ML detection are also shown.

\begin{figure}[t]
	\centerline{\includegraphics[width=3.65in]{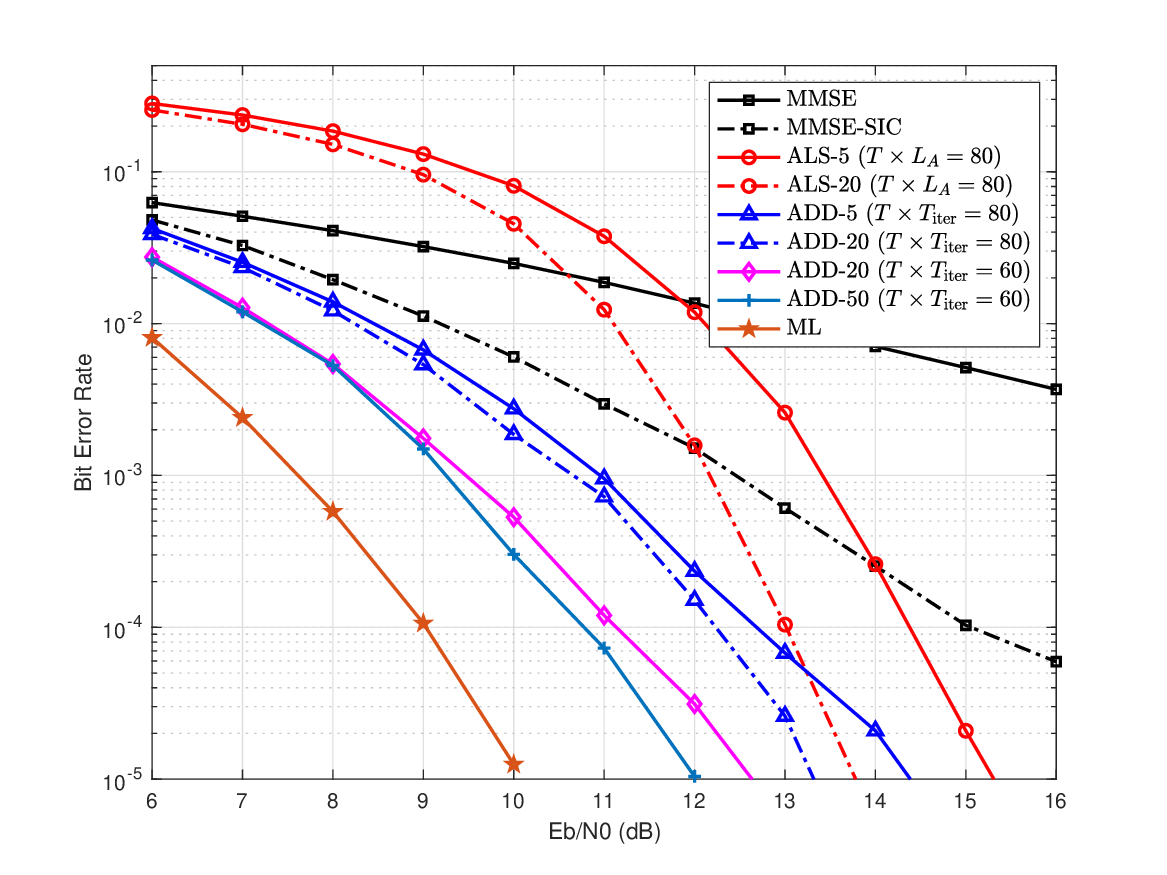}}
	\vspace{-.5em}
	\caption{Performance comparison between the proposed ADD and ALS method for $32 \times 32$ system using $4$-QAM.}
	\label{fig:performance}
	\vspace{-1em}
\end{figure}

Fig.\ref{fig:performance} shows the bit error rate (BER) for an uncoded system with $N = K = 32$ and $4$-QAM.
For fair comparison, we first consider $T \times L_A = T \times T_\mathrm{iter}$ case, where ADD and ALS can be taken as implementing for the same iterations.
Clearly, with 5 calls, the performance of ADD is more satisfying than that of ALS, and by increasing the sampling trajectories, further performance improvement can be observed by both of the sampling detector.
The reason about this gap between ADD and ALS might be the intrinsic property of SNIPS method, where the diffusion noise is assumed in a special way to promise the dependence on the channel noise, but this assumption might be less reliable in the low signal-noise-ratio region. 
An interesting thing about the ADD is, that the performance can be even enhanced with a fewer $T_\mathrm{iter}$ (See ADD-20, $T_\mathrm{iter} = 3$).
This is similar to the phenomenon found in GPGD scheme\cite{GPGD}, claiming that there exists the most suitable $T_\mathrm{iter}$ for particular linear iterative method.
Besides, after increasing the trajectories of ADD to $50$, only a small gain can be observed, which implies that the convergence of ADD is rather fast but its performance upper bound still gets a nearly $2$dB gap to the ML detection.
Recall that this ADD instance just adopts a simple linear iterative detector, which could have eventually converged to the linear MMSE detection.
Nevertheless, it still manages to significantly outperform the nonlinear MMSE-SIC and partially recover the gap to ML detection.

\begin{figure}[t]
	\centerline{\includegraphics[width=3.65in]{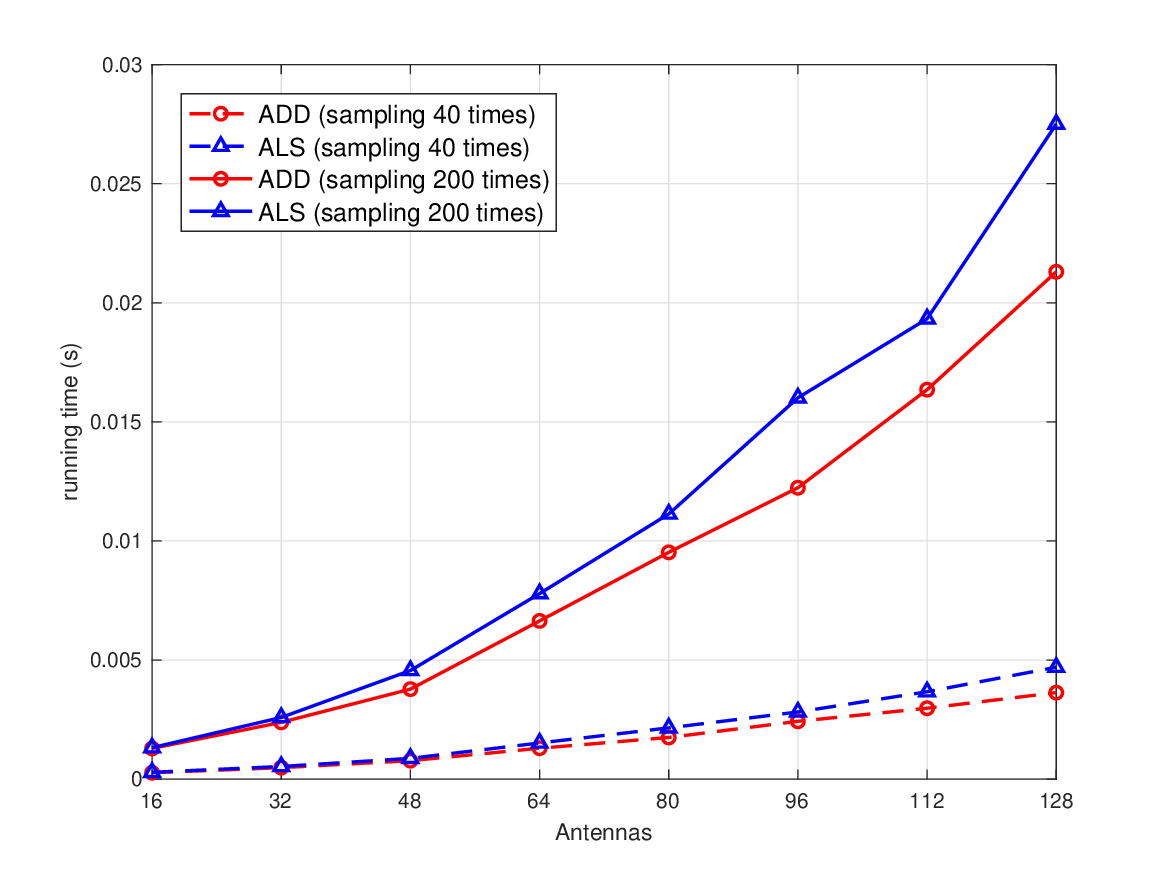}}
		\vspace{-.5em}
	\caption{Running time of ADD and ALS for one trajectory ($N = K$).}
	\label{fig:running time}
	\vspace{-1em}
\end{figure}

Fig.\ref{fig:running time} compares the average running time of ADD and ALS for one single trajectory.
The ADD is configured as $T_\mathrm{iter} = 1$.
It can be seen that these two methods show a similar increasing tendency with the dimension, while the ADD still slightly enjoys a more competitive time complexity for the absence of SVD operation.
Besides, as shown in Fig.\ref{fig:performance}, the ADD outperforms ALS under the same configuration, especially at low signal-noise-ratio regime, thereby achieving a better complexity-performance trade-off.

\section{Conclusion}
In this paper, we proposed an approximate diffusion detection (ADD) strategy for massive MIMO systems, which can be applied to a wide range of iterative detectors and eliminates the SVD in existing score-based method.
The resultant ADD turned a deterministic iterative detector into a stochastic one to achieve performance gain through sampling.
This introduced inner iterative detector helped ADD produce samples based on an approximate posterior, thereby circumventing the SVD required for explicit score calculation.
The conducted simulations verified that a fraction of the gap to ML detection can be recovered by the adoption of ADD scheme.
Besides, the presented ADD customized by a CGD detector enjoys a more satisfying complexity-performance trade-off compared to the existing score-based detector, but more attempts on this replaceable detector are encouraged.
Besides, we discussed the possibility of extending the ADD scheme to a deep generative detection network, which constitutes our future work.


\section{Acknowledgment}
This work was supported in part by National Natural Science Foundation of China under Grants No. 62371124, and in part by the National Key R\&D Program of China under Grants No. 2023YFC2205501.

\end{document}